\documentclass[]{spie}
\usepackage[utf8]{inputenc}

\newcommand{\ocs}{\texttt{ocs}}
\newcommand{\soaculib}{\texttt{soaculib}}
 
\usepackage{amsmath,amsfonts,amssymb}
\usepackage{graphicx}
\usepackage[colorlinks=true, allcolors=blue]{hyperref}
\usepackage{lineno}
\usepackage{gensymb}
\usepackage{xcolor}
\usepackage{longtable}

\title{The Simons Observatory: Antenna control software integration and implementation}

\author[a]{Lauren J. Saunders}
\affil[a]{Department of Physics, Yale University, New Haven, CT 06511, USA}
\author[b]{Matthew Hasselfield}
\affil[b]{Center for Computational Astrophysics, Flatiron Institute, New York, NY 10010, USA}
\author[a]{Brian J. Koopman}
\author[a]{Laura Newburgh}

\authorinfo{Corresponding author: \href{mailto:lauren.saunders@yale.edu}{lauren.saunders@yale.edu}}
\begin{document}

\maketitle

\abstract
The Simons Observatory (SO) is a ground-based cosmic microwave background survey experiment that consists of three 0.5 m small-aperture telescopes and one 6 m large-aperture telescope, sited at an elevation of 5200 m in the Atacama Desert in Chile. SO will study the polarization and temperature anisotropies of the Cosmic Microwave Background (CMB). The observatory will require well-understood telescope pointing and scanning. Good antenna control will allow us to execute the scan strategy devised to optimize sensitivity to our scientific goals, calibrate the system with celestial targets, and make maps. To achieve this, we integrate the data acquisition and control of the telescopes’ Antenna Control Units (ACUs) within the software framework of the SO Observatory Control System (OCS). We present here the current status of the software integration for the ACUs, as well as measurements of the Small Aperture Telescope platforms’ responsiveness to software commanding in the factory, plans for in situ measurements, and prospects for implementation on the Large Aperture Telescope.
\endabstract{}


\section{Introduction}\label{sec:intro}
The Simons Observatory (SO) is a new ground-based Cosmic Microwave Background (CMB) experiment being constructed in the Atacama Desert in Chile. The observatory will consist of three 0.5 m Small Aperture Telescopes (SATs)\cite{2020JLTP..200..461A} and one 6 m Large Aperture Telescope (LAT)\cite{2021RNAAS...5..100X}. The combination of Large and Small Aperture Telescopes will cover a wide range of science goals, including constraining the sum of the neutrino masses and effective number of relativistic species, measuring or constraining the primordial tensor-to-scalar ratio, and measuring the duration of reionization\cite{simons_2018, https://doi.org/10.48550/arxiv.1907.08284}. To accomplish these goals, the observatory will field more than 60,000 cryogenic bolometers across the four telescopes, and will also require a variety of instruments to modulate the sky signal, maintain the operations of the observatory, and monitor the operating conditions of each telescope.

Conducting observations with any observatory requires well-organized data acquisition, control, and monitoring for all of these instruments. Observations with motorized telescopes additionally requires adherence to scientifically-motivated scan strategies, and coordinated responses from a series of systems within the telescope to take detector data, perform ancillary maintenance activities, and acquire auxiliary data. While such data acquisition and control software systems exist on other telescopes\cite{Story_2012}, these systems do not meet the requirements for use on SO. To ensure that the SO data acquisition, control, and monitoring needs are met, we have designed the Observatory Control System (\ocs)\cite{10.1117/12.2561771}.

In these proceedings, we present an overview of the software used to operate the telescope platforms for the Simons Observatory. In Section \ref{sec:ocs}, we give a brief overview of the \ocs{} software. In Section \ref{sec:acu}, we discuss the capabilities of the Antenna Control Unit (ACU) software, which provides the interface for data acquisition and control of the telescope platforms. In Section \ref{sec:acu_agent}, we detail the ACU Agent, which is the interface between the ACU software and \ocs. In Section \ref{sec:fat}, we discuss software testing completed with the SAT platforms in the factory setting. In Section \ref{sec:future}, we discuss future work with the ACU and \ocs{} software, including expected updates for the LAT and planned testing with the SATs in situ. Appendix \ref{sec:appendix} contains tables of data fields collected via the ACU Agent.

%%%%%%%%%%%%%%%%%%%%%%%%%%%%%%%%%%%%%%%%%%%%%%%%%%%%%%

\section{Observatory Control System} \label{sec:ocs}
The Observatory Control System (\ocs)\footnote{\url{https://github.com/simonsobs/ocs}} is a distributed data acquisition, control, and monitoring system. It is designed for ease of use in the complex, distributed systems commonly seen in modern observatories. \ocs{} is described in detail in Ref.~\citenum{10.1117/12.2561771}; it is described briefly here for context.

The \ocs{} architecture provides two main components: Agents and Clients. \ocs{} Agents are servers that act as the interface between the larger \ocs{} framework and a hardware or software component of the observatory. Each Agent contains operations denoted as Tasks and Processes. A Task is an operation that completes a job and is expected to do so in a finite amount of time; for example, commanding the telescope platform to move from one position to another would be a Task. Processes run continuously until commanded to stop, and are frequently used for continuous data acquisition. \ocs{} Clients are objects that call the operations allowed by the Agent. \ocs{} Agents communicate via a Publish/Subscribe (PubSub) pattern, and Tasks and Processes may run on Agent startup or be called by a Client by a Remote Procedure Call (RPC). These communications are facilitated by a central Web Application Messaging Protocol (WAMP) router; \ocs{} uses \texttt{crossbar.io}\footnote{\url{https://crossbar.io}} as its implementation of this router. 

Data collection and live monitoring are handled separately for detectors and all other hardware objects. The details of the detector data acquisition and monitoring are beyond the scope of these proceedings. For all other devices, which we refer to as ``housekeeping'' devices, data collected by \ocs{} Agents are published to \ocs{} Feeds, which are a communication layer on top of the basic PubSub functionality used to pass data among different Agents and Clients. Due to differences in the individual housekeeping components, the data within these feeds are published asynchronously at varying rates. The Housekeeping (HK) Aggregator Agent subscribes to these Feeds and writes these data to disk in the G3 file format.

To conduct remote live-monitoring of the housekeeping data sets, we use Grafana\footnote{\url{https://grafana.com/}}, an open-source web-based application designed for visualization and analytics of time-series data. Data is shown in plots called ``panels'', accessible on web pages called ``dashboards''. Grafana allows a configurable data backend; for the \ocs, we use InfluxDB, a popular data source backend developed by InfluxData\footnote{\url{https://www.influxdata.com/}}. InfluxDB is designed for quick access to time-series data, and is suitable for loading data for most SO housekeeping data needs.

We additionally implement a custom \ocs{} web interface, which is written in JavaScript and rendered as a GUI in a web browser. \ocs{} web provides control panels and data feed values for all Agents connected to \ocs.

%%%%%%%%%%%%%%%%%%%%%%%%%%%%%%%%%%%%%%%%%%%%%%%%%%%%%%

\section{Telescope Platform and Antenna Control Unit} \label{sec:acu}
We discuss here the telescope platforms, which are the hardware frameworks that support the telescope receivers, as well as the Antenna Control Unit, a computer used to communicate with and control the platforms. The physical platforms for the SATs and LAT and the ACU are developed by Vertex Antennentechnik GmbH\footnote{ \url{https://www.vertexant.com/}} (referred to as ``VA'' throughout these proceedings).

\subsection{Telescope Platforms}\label{subsec:platforms}
\begin{figure} [h!]
\begin{center}
\begin{tabular}{c}
\includegraphics[width=0.6\linewidth]{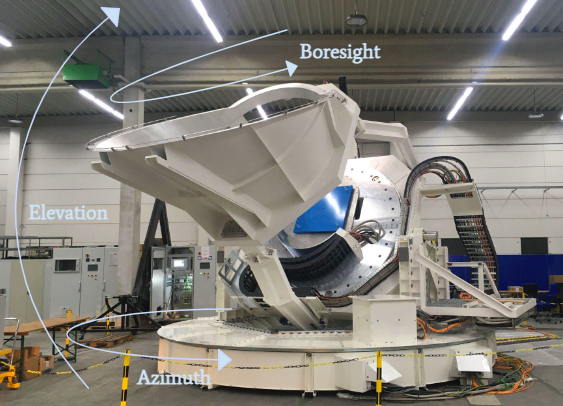}
\end{tabular}
\end{center}
\caption[example]
{ \label{fig:axis_labels}
The three SAT axes of rotation are depicted with a picture of a SAT for reference. The Azimuth axis corresponds to the spherical azimuth angle, the Elevation axis to the complement of the spherical polar angle, and the Boresight axis to the rotation about the center of the focal plane.
}
\end{figure}

The platforms for the SATs and LAT support the telescope receivers and the LAT mirror. All four platforms are constructed by VA, and are fully assembled and tested at the VA company facility in Duisburg, Germany. Each platform has three axes of rotation: the Azimuth, Elevation, and Third axes (see Figure \ref{fig:axis_labels}). The Azimuth axis refers to the spherical azimuth angle; each telescope platform may traverse 570\degree{} of rotation on this axis. The Elevation axis refers to the complement of the spherical polar angle, with allowed angles between 20\degree{} and 90\degree. The Third axis refers to the rotation of the receiver; on the SATs, the Third axis is referred to as the Boresight axis. Motors power the motion of each axis; for the SATs, the Azimuth and Boresight rotations are powered by two motors for each axis, while the Elevation rotation is powered by one motor. Encoders along each platform axis measure rotation, and are interpreted by the Antenna Control Unit (ACU). Each axis is additionally equipped with brakes to arrest motion, and limit switches to prevent rotations beyond the allowed motion limits.

\subsection{ACU Overview}\label{subsec:acu}
The ACU is an industrial computer with specialized device cards and VxWorks, a real-time operating system, installed. It is housed in an electronics rack and supports a wired connection to the telescope platform. The ACU software and interfaces were developed by VA, in consultation with SO representatives. The ACU is used for both data collection and interpretation, and for controlling the telescope motors. It provides a local interface for data visualization and commanding, as well as a remote interface. Users can connect to the ACU remotely via HTTP interface, and can collect data from this HTTP interface as well as via a User Datagram Protocol (UDP) server. The data and commanding available through these connections is discussed in Section \ref{subsec:acu_software}.

The ACU additionally has a built-in Emulator Mode to enable software development. This operational mode provides a software emulator that is programmed to behave like an ideal telescope platform. The ACU interprets the emulated encoder positions and allows users to perform the allowed commands in Emulator Mode. The development of software described in Section \ref{sec:acu_agent} was completed with the ACU in Emulator Mode; following the completion of this software development, we performed initial tests with the SAT platforms, described in Section \ref{sec:fat}.

\subsection{ACU Software Overview}\label{subsec:acu_software}
The ACU software is built by VA in conjunction with the telescope platform hardware, and is updated as needed to accommodate SO data collection and commanding needs, correct software bugs, and make routine updates. To ensure accurate timestamps, the ACU's internal clock is synchronized to a GPS signal using Precision Time Protocol (PTP). This timestamp is included in all data from the ACU.

The ACU outputs a 200 Hz data stream via UDP. For the SATs, this stream contains timestamps, encoder positions for each axis, and the currents through the motors on each axis. Each axis also has two recorded encoder positions: the true encoder position, and the encoder position after any pointing model corrections applied by the ACU. During 48 hours of continuous data collection from this stream, we found no dropped data frames.

The ACU HTTP server provides several data sets at 20 Hz that can be accessed via web browser or the command line. These include data sets containing details about the pointing model and refraction corrections, which can be set by the user; parameters specific to the third axis; faults and errors; and the general status of the platform. The values that are most frequently needed to understand the performance of the SAT platform are also available under a Detailed SAT Platform (SATP) Dataset. These values include the pointing model-corrected encoder positions for each axis; booleans for whether limits, faults, and errors are triggered; and general status information about the platform and the ACU itself.

The ACU software contains an internal parameter to control the motion mode, which informs the software of how to calculate the telescope trajectory. There are several available motion modes, including: stop mode, which engages the brakes and prevents current from going through the motors; a point-to-point motion mode, which commands the platform to move to a position and maintain that position; and an uploaded scan mode, which directs the telescope to follow a path that is uploaded to the ACU as points by the user.

This HTTP server additionally provides the interface for sending commands via the ACU. Motion types, which use the above modes, include: point-to-point motions at the maximum allowed velocity; stopping, which engages the brakes; constant-velocity motion, where the velocity is set by the user; star tracking; stow mode, which moves the telescope to a pre-chosen position when not in use; and scan tracking, which traces a path between points in time and position that are uploaded by a user. The uploaded scan requirements and features are discussed in the next paragraph.  Other commands include clearing any uploaded scan points from the stack, changing the motion mode of an axis, and changing pointing model and refraction parameters. Motion commands require a change in the telescope motion Mode, which allows the ACU to correctly interpret motion types.

The ACU software is designed to calculate trajectories between points for a scan to best create a smooth, well-controlled motion. To execute a scan, the user may upload a text file to the ACU via a scan points upload HTTP interface. Each line of the file must contain seven data points: the time, in the format \texttt{Day of Year, Hour:Minute:Second}; the values of Azimuth and Elevation to reach at those times; the velocities of the motion along each of these axes; and flags for each type of axis motion, which denote whether the point refers to a position within the constant-velocity part of a scan, is the final point before a turnaround, or is in an undefined part of the scan. The uploaded times must be spaced a minimum of 0.05 seconds apart. A maximum of 10,000 points may be uploaded to the stack at any time, and points may be added to the stack whenever there are free positions; this allows scans of any length to be uploaded in batches of $\leq$10,000 points. As an hour-long scan would require 72,000 uploaded points, we preference the method of uploading a smaller number of points at $\sim$1 minute intervals. Using these uploaded points, the ACU calculates the trajectory. When the motion is a constant-velocity portion of a scan, the ACU calculates the trajectory with a linear interpolation. For turnarounds and undefined scans, it instead applies a cubic spline interpolation.

%%%%%%%%%%%%%%%%%%%%%%%%%%%%%%%%%%%%%%%%%%%%%%%%%%%%

\section{ACU Agent} \label{sec:acu_agent}
The ACU \ocs{} Agent is the software interface between the ACU software and the \ocs{} framework. It facilitates all of the data acquisition, live monitoring, and commanding for the ACU for the SO \ocs{} user, and enables the ACU to be used in conjunction with other telescope systems when conducting observations. In this section, we discuss the custom Python library used with the ACU Agent, called \soaculib{} (Section \ref{subsect:soaculib}), Agent operations (Sections \ref{subsect:daq}, \ref{subsect:livemonitor}, and \ref{subsect:command}), and a software simulator of the ACU (Section \ref{subsect:sim}).

\subsection{\soaculib{}}\label{subsect:soaculib}
\soaculib{}\footnote{\url{https://github.com/simonsobs/soaculib}} is a Python library and function repository custom-built to support ACU Agent communications with an ACU. While it was built with this goal in mind, \soaculib{} may also be used for communications with the ACU outside of the \ocs{} framework. The \soaculib{} library includes HTTP protocol setup, the ACU-specific HTTP statements, configurations for each ACU, and information about the fields within the Detailed SATP Status Dataset.

\soaculib{} provides an HTTP backend interface to streamline the HTTP protocol usage within the ACU Agent. We provide options for a backend using the Python \texttt{requests} library (which we refer to as the ``standard backend'') or a backend using an instance of the Python \texttt{twisted} library (which we refer to as the ``twisted'' backend). We most commonly use the standard backend for debugging HTTP communications outside of \ocs{} testing, and use the twisted backend for \ocs{} communications, as this library allows us to perform asynchronous requests. The library additionally contains the vocabulary necessary to request values from and send command requests to the ACU. To ensure that the Agent queries the correct ACU Datasets and receives the expected value fields, the \soaculib{} library also contains configuration files for each ACU that may be connected to the \ocs{} computer, as well as a dictionary of the field keys expected from the primary Status Dataset queried by the Agent.  This field key dictionary additionally assigns short, descriptive names to each field using only characters allowed in the \ocs{} Feed field names.

\subsection{Data Acquisition Processes}\label{subsect:daq}
As detailed in Section \ref{subsec:acu_software}, the ACU streams a small subset of data fields at 200 Hz via UDP, and additionally supports a wide array of data fields that may be queried via HTTP connection, divided into several Datasets. For SO data acquisition purposes, we collect both the 200 Hz data and the 20 Hz Detailed SATP Status Dataset, managed by Agent Processes \texttt{broadcast} and \texttt{monitor}, respectively.

The \texttt{broadcast} Process sets up a UDP server to collect all packets broadcast via UDP on a designated port. Each packet contains 10 data points, which are in a binary Datagram format. The \texttt{broadcast} function accesses the binary format as it is stored in the configuration file within \soaculib{}. It uses the \texttt{struct} module in Python to calculate the length of each datum and decode the data. Data are then added to a list within the function, and processed for publishing to the OCS Feed in groups of 200. For each datum within the group, we use the ACU-generated day of year and second of day timestamps to calculate the time since the epoch (\texttt{ctime}), so that we record a unique timestamp for each datum. The other data fields produced by the ACU in this data stream are additionally collected and separated by index. Using the data field names, as stored in the \soaculib{} configuration file, the \texttt{broadcast} function builds a dictionary of each field name and corresponding datum. The dictionary is then published to an OCS Feed. The HK Aggregator Agent subscribes to this feed and records the data on disk in G3 files.

The \texttt{monitor} Process utilizes the tools from \soaculib{} to query the HTTP server for the Detailed SATP Dataset. Data from this Dataset is then divided into blocks, which are structured according to the \texttt{status\_keys} module of \soaculib{} and renamed according to the key dictionary of this same module. \texttt{monitor} then publishes the blocks to one OCS Feed, giving each block a descriptive block name. The HK Aggregator Agent subscribes to this feed and records the data on disk. All data types output by the ACU Detailed SATP Dataset (strings, integers, floats, Nones, and booleans) are accepted by the HK Aggregator Agent and saved to G3 files.

The HK data fields are described in Appendix \ref{sec:appendix}. Both \texttt{broadcast} and \texttt{monitor} have additional features to handle data recorded with InfluxDB, which is described in Section \ref{subsect:livemonitor}.

\subsection{Live Monitoring}\label{subsect:livemonitor}
While InfluxDB is well suited as a live monitoring backend for most data, it can experience a substantial slowdown when large amounts of data are queried. The 200 Hz \texttt{broadcast} data from the ACU Agent runs the risk of causing a significant slowdown of the backend server when querying the data for live monitoring. To reduce this load, we have chosen to publish only every tenth datum from this data set to InfluxDB, such that the data rate as recorded by InfluxDB is 20 Hz. The \texttt{broadcast} Process of the ACU Agent therefore publishes data to a second OCS Feed, which contains only these points and is not saved to disk in the G3 file format.

Within the Detailed SATP Dataset, most values are either floats or integers. However, the motion modes are written as strings, which may be published as-is to an OCS Feed, but cannot be interpreted by InfluxDB. To accommodate live monitoring for the motion mode fields, we assign each string an integer value: for example, stop mode is assigned to value 0, point-to-point mode to 1, and uploaded scan mode to 2. We additionally translate all boolean values to their 0 and 1 integer counterparts, and publish these translated values to an OCS Feed that is subscribed to only by the InfluxDB Publisher Agent. In addition, the Detailed SATP Dataset includes values for commanded positions on each axis; this value is set to None when no position is commanded. Because the None type is not accepted by InfluxDB, we write to a special OCS Feed that only publishes these values to InfluxDB when the value is not None.

\subsection{Command Operations}\label{subsect:command}
The ACU Agent is written to support six operations for sending commands to the ACU. These operations are: \texttt{go\_to}, \texttt{set\_boresight}, \texttt{scan\_fromfile}, \texttt{constant\_velocity\_scan}, \texttt{generate\_scan}, and \texttt{stop\_and\_clear}.

\begin{figure} [h!]
\begin{center}
\begin{tabular}{c} 
\includegraphics[width=0.4\linewidth]{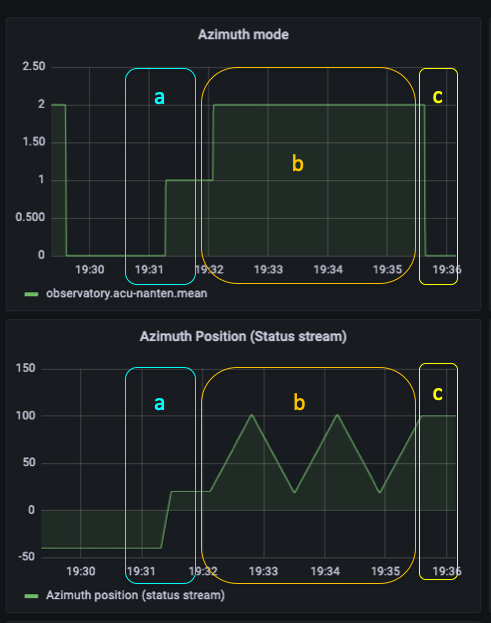}
\end{tabular}
\end{center}
\caption[example]
{ \label{fig:motion_types}
Three motion types are executed on the Azimuth axis in sequence by an ACU in Emulator Mode, shown here in a Grafana dashboard set to monitor the ACU. The top panel shows the Azimuth mode as translated by the Agent for InfluxDB (where mode 0 is stop mode, mode 1 is point-to-point motion mode, and mode 2 is uploaded scan mode). The bottom panel shows the Azimuth position. In this motion, the motion mode is changed three times. In segment a, the ACU is switched to point-to-point motion mode, and completes a motion at maximum speed from -40\degree to 20\degree. In segment b, the ACU is switched to the uploaded scan mode and completes a scan with velocity $\pm$2\degree/second, scanning back and forth between 20\degree and 100\degree. Finally, in segment c, the ACU is switched to stop mode, the brakes are engaged, and all motion on the Azimuth axis is stopped.
}
\end{figure}

The \texttt{go\_to} Task is used to set specified Azimuth and Elevation positions for the telescope platform. This Task has two required input values, the desired Azimuth and Elevation values; as well as three optional inputs: the time for the Task to wait for motion to begin (default value 0.1 seconds), the rounding precision for determining the platform's current position (default value 1 decimal place), and a parameter for activating the brakes and setting all three axis Modes to Stop at the end of the motion (the default is to not perform this stop). This function therefore has two purposes: it can be used to set new positions in Azimuth and Elevation and move to them, or it may be used to set the Azimuth and Elevation axis motion modes to the point-to-point mode and maintain the current position. For both uses, the Task first checks the current position, then calls the \texttt{go\_to} method of \soaculib{}. If the input Azimuth and Elevation positions are the same as the current position within the specified rounding precision, the Task completes after calling the \soaculib{} \texttt{go\_to} method. When moving to a new position, the Task first checks whether the platform has begun moving by reading the velocity values and waiting for them to change from 0.0\degree/s. If the velocities stay at 0.0\degree/second for longer than 30 seconds, the value for the current position is checked. In the case that the current position has not changed from the initial position, the Task is aborted, the brakes are engaged, and the telescope is set to Stop mode to investigate why the platform has not moved. If the axis velocities do change, the velocity values are monitored until they are both close to 0.0\degree/second, within the degree of rounding precision specified when calling the Task. The Task then monitors the Azimuth and Elevation position values, and waits until these values are within the specified rounding precision. If the parameter to stop the platform after the motion is set to True, the Task then calls the \soaculib{} \texttt{stop} command, which engages the brakes and sets all three axis Modes to Stop. An example of this Task is shown in Figure \ref{fig:motion_types}.

The \texttt{set\_boresight} Task is similar to the \texttt{go\_to} Task, but is used to set the value for the Boresight or 3rd Axis only. Because the Boresight axis is controlled separately from the other two axes by the ACU, this Task uses the \texttt{go\_to\_3rd\_axis} method of \soaculib{}, but otherwise follows the logic of the \texttt{go\_to} Task.

Our ability to command the platforms to complete scans is vital to our planned observation strategies. Most planned scans will involve constant-velocity Azimuth rotations at a set Elevation. To command these scans, we implement the \texttt{constant\_velocity\_scan} Task within the ACU Agent. This Task calls a specialized helper function that uses input parameters for the Azimuth scan endpoints, Elevation value, Azimuth scan velocity, and the Azimuth turnaround acceleration, to calculate a scan trajectory. This trajectory includes ordered lists of times to reach each position, Azimuth values, Elevation values, velocities on each axis, and ACU scan flags for each axis. Although the points generated with this function are evenly spaced along the constant-velocity portions of the scan, the function uses the input acceleration to calculate a time gap for the turnaround. We leave this gap for interpretation by the ACU cubic spline interpolation of the turnaround. Another helper function then writes this information as a list of strings in the format \texttt{Day of Year, Hour:Minute:Second;} \texttt{Azimuth Position,} \texttt{Elevation Position,} \texttt{Azimuth Velocity,} \texttt{Elevation Velocity,} \texttt{Azimuth Flag,} \texttt{Elevation Flag}. Alternatively, for scans planned with variable Azimuth velocities or varying Elevation positions, we can create a \texttt{numpy} file containing the ordered lists for time, Azimuth, Elevation, Azimuth velocity, Elevation velocity, Azimuth flags, and Elevation flags. The \texttt{fromfile\_scan} Task of the ACU Agent calls helper functions that read the \texttt{numpy} file and reformat the information into the list of strings in the correct format. Both of these Tasks then call an ACU Agent function that changes the axis Modes to PtStack and uploads the lines to the ACU in batches using the \soaculib{} \texttt{PtStackUpload} method. While the scan is running, this function monitors the number of free upload stack positions; when a certain number of positions are free, it uploads another batch of lines, until there are no more lines to upload. It then monitors positions and velocities to determine when the scan has concluded.

The length of these constant-velocity scans is limited by the control computer memory, as all of the upload lines must be stored in RAM. This causes significant slowdowns and hanging when scans last more than a few hours. For arbitrarily long scans, we implement another operation, \texttt{generate\_scan}. This Process calls a helper function which establishes a Python generator to create new upload points in batches. These points are then converted to the necessary line format and uploaded to the ACU in batches in the same way as for the other scan types. The upload frequency is determined by the number of free upload stack positions, such that the number of available upload points rarely exceeds 5100. The \texttt{generate\_scan} function is only used for constant-velocity Azimuth scans.

\begin{figure} [h!]
\begin{center}
\begin{tabular}{c} 
\includegraphics[width=0.8\linewidth]{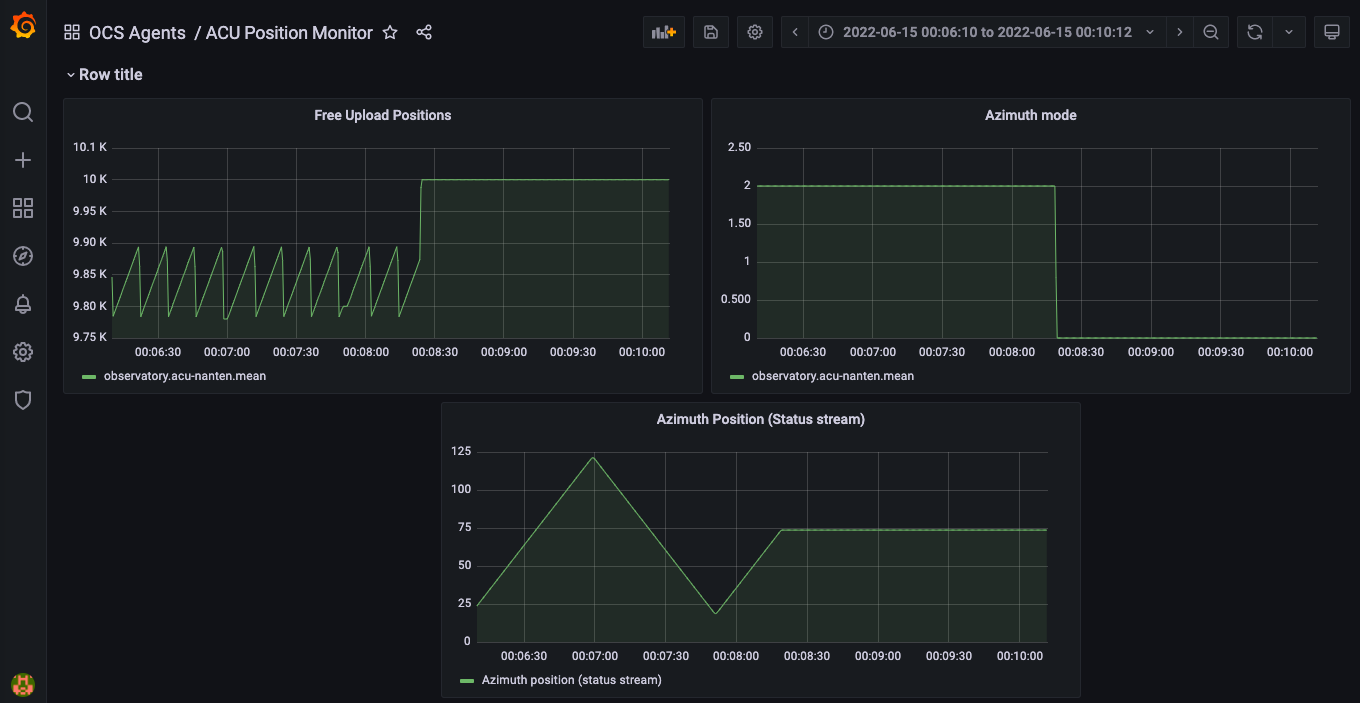}
\end{tabular}
\end{center}
\caption[example]
{ \label{fig:stopclear}
While a constant-velocity scan is proceeding, the user runs a \texttt{stop\_and\_clear} Task via an \ocs{} Client, shown in the Grafana dashboard set up to monitor an ACU in Emulator Mode. We show the number of free upload positions (upper left), the Azimuth motion mode (upper right), and the Azimuth position (bottom). The \texttt{stop\_and\_clear} Task changes the motion mode from the upload scan mode (interpreted as 2) to stop (interpreted as 0). The brakes are applied, so all motion on the Azimuth axis is ended. The Task then clears all of the uploaded points from the stack, resetting the number of free stack positions to 10000. The operation shown in this figure was performed with an ACU in Emulator Mode.
}
\end{figure}

The operation \texttt{stop\_and\_clear} is a Task whose primary purpose is to activate the platform brakes and clear out any values that may have been uploaded to the stack for a scan. The platform brakes are activated with an \soaculib{} \texttt{stop} function. This sends a stop command to the ACU and resets the Mode for all telescope axes to Stop. The stack is cleared using an \soaculib{} \texttt{Command} function, with parameter `Clear Stack'. The \texttt{stop\_and\_clear} Task can be used to abort other motion operations. An example of the \texttt{stop\_and\_clear} Task being used to abort a scan is shown in Figure \ref{fig:stopclear}.

\subsection{ACU Simulator}\label{subsect:sim}
Building a full-observatory testing environment is essential to testing the software stability and user control interface of the ACU Agent, as well as the full \ocs{} software system. Due to limited availability of the ACU and the need to test with three SATs and one LAT together, these tests require a piece of software that can simulate the behavior of an ACU without needing to access the actual ACU in Emulator Mode. It is important that this simulator correctly handles commands from the ACU Agent, and generates data identically to the real ACU, so that it can be used with the ACU Agent and data can be interpreted by other pipeline software.

The ACU simulator consists of a multi-threaded server that stores and updates simulated ACU data, and serves this data via a UDP server and an HTTP server. The data is generated, stored, and updated with a single Python class, the UDP server uses the standard \texttt{socket} module, and the webserver is built using the Flask\footnote{\url{https://flask.palletsprojects.com/}} framework. Data may be queried from the HTTP server, and commands can be sent using the same protocols used by \soaculib{}. The ACU simulator can be used with the ACU Agent to simulate data output, \texttt{go\_to} and \texttt{set\_boresight} commanding, \texttt{fromfile\_scan} and \texttt{constant\_velocity\_scan} Tasks, and the \texttt{generate\_scan} Process.

%%%%%%%%%%%%%%%%%%%%%%%%%%%%%%%%%%%%%%%%%%%%%%

\section{SAT Factory Testing} \label{sec:fat}
We performed a series of tests to ensure correct ACU functioning before the ACUs and SAT platforms were shipped to Chile. These tests were performed with the ACUs connected to SAT platforms; the ACUs were additionally connected to a computer with the \ocs{} and \soaculib{} software installed. \ocs{} users connected remotely via SSH to the \ocs{} computer. Tests were divided into two sections: data acquisition and trigger testing, and \ocs{} command testing. Data verification and analysis was completed offline.
\subsection{Data Acquisition Testing} \label{subsec:fat_daq}
Data acquisition testing was performed simultaneously with a series of VA-administered tests. With the ACU Agent and \ocs{} data aggregation and live monitoring running, a VA technician stepped through a variety of procedures to trigger each expected limit, fault, error, and warning within the Detailed SATP Status Dataset. This allowed us to exercise the interface and ensure that the Detailed SATP Status Dataset was fully compatible with the SO data requirements and the ACU Agent. All discrepancies were corrected, so that for each triggered value, the \ocs{} users verified that the value was also triggered in the Dataset as seen on the \ocs{} web interface.

\subsection{Command Testing} \label{subsec:fat_cmd}
The initial set of command tests were performed as part of the VA-administered testing. To ensure that all aspects of remote commanding of the ACU function as expected, we uploaded new refraction and pointing model parameters to the ACU and monitored the response in the Datasets. We observed that, when left in point-to-point motion Mode, the platform axis positions changed slightly to account for these new parameters, as expected.

\begin{figure} [h!]
\begin{center}
\begin{tabular}{c}
\includegraphics[width=0.7\linewidth]{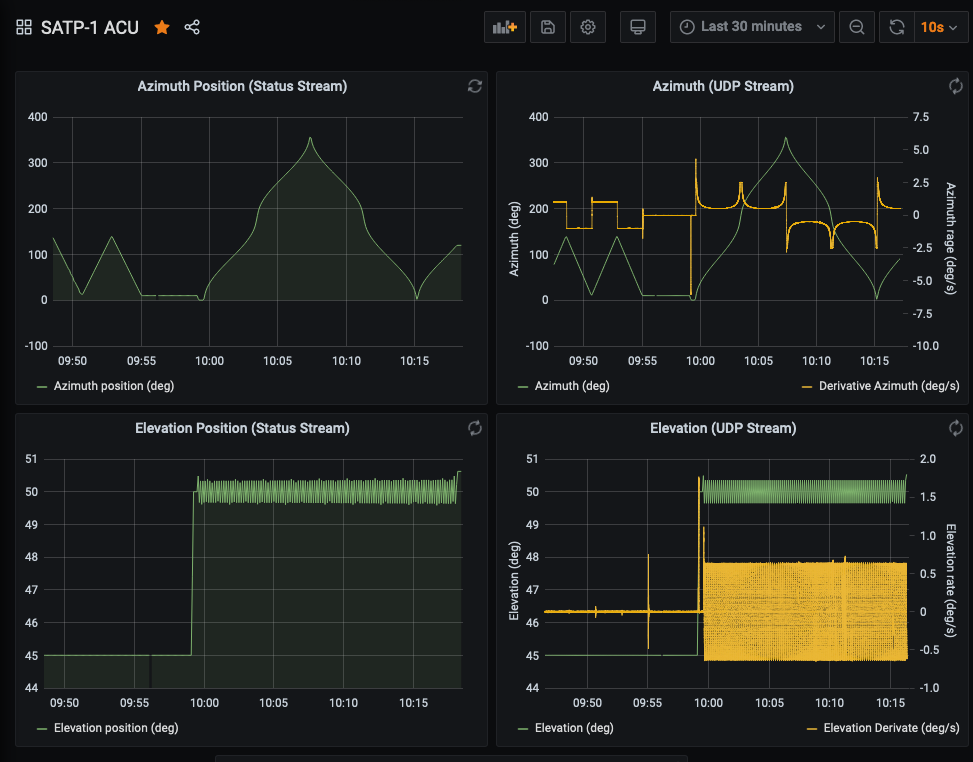}
\end{tabular}
\end{center}
\caption[example]
{ \label{fig:fromfilescan}
Screenshot of the Grafana Dashboard during a \texttt{fromfile\_scan} Task. The scan executed here involves a varying-velocity Azimuth motion and a sinusoidal Elevation motion. The top left panel shows the Azimuth position in green as recorded with the \texttt{monitor} process. The top right panel shows the Azimuth position in green as recorded by the \texttt{broadcast} process, along with a time derivative calculated by Grafana in yellow. The bottom left panel shows the Elevation position in green as recorded by the \texttt{monitor} process. The bottom right panel shows the Elevation position in green as recorded by the \texttt{broadcast} process, along with a time derivative calculated by Grafana in yellow.
}
\end{figure}

Following the VA-administered software testing, we performed a series of tests to demonstrate \ocs{} control of the telescope platforms via the ACU, supervised by VA. We began by performing a series of commands to ensure functionality: we performed a point-to-point motion with the \texttt{go\_to} Task, a short scan with the \texttt{constant\_velocity\_scan} Task, and a Boresight rotation with the \texttt{set\_boresight} Task. The \texttt{stop\_and\_clear} Task was executed between each motion. During the testing with the SAT 3 platform, we additionally executed a \texttt{fromfile\_scan} with varying Azimuth velocity and a commanded Elevation sinusoidal motion (Figure \ref{fig:fromfilescan}). The trajectory was executed as expected by the ACU.

Several further tests were performed to better understand ACU and telescope platform motion performance, particularly during and at the end of scans. To perform these tests, we ran several \texttt{constant\_velocity\_scan} Tasks with varying velocities and turnaround accelerations. These tests are discussed in Section \ref{subsec:fat_analysis}.

\subsection{Data Verification}
\label{subsec:fat_analysis}

Because constant Elevation and constant Azimuth velocity scans will be our primary scanning pattern, we collected data during the factory testing with the SATs to best understand ACU and platform behavior during these types of motions.

\begin{figure} [h!]
\begin{center}
\begin{tabular}{c} 
\includegraphics[width=0.6\linewidth]{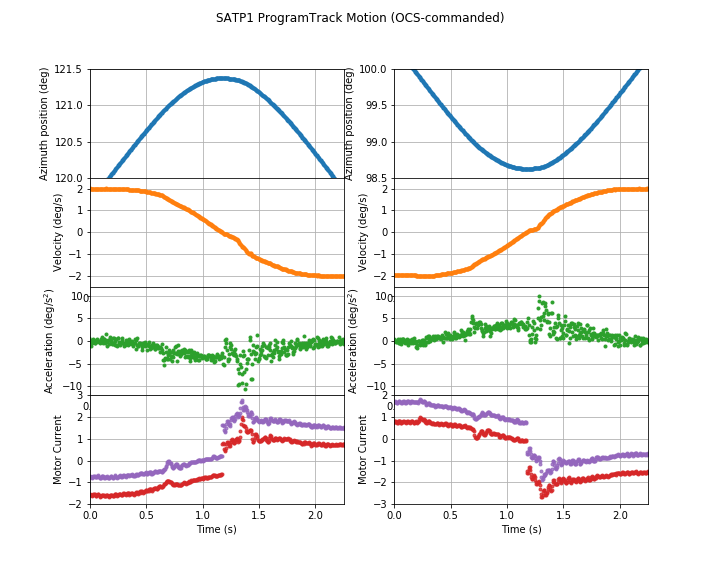}
\end{tabular}
\end{center}
\caption[example]
{ \label{fig:progtrack_currs}
An example of turnarounds in an \ocs-controlled Azimuth scan, from positive to negative velocity (left) and negative to positive velocity (right). The plots in each column, from top to bottom, show the Azimuth position, velocity, acceleration, and motor currents. The brief flattening of the velocity curves, as well as the peaks in acceleration and motor currents, are expected from the platform design.}
\end{figure}

An important consideration in executing these scans is the smoothness and acceleration during turnarounds. Due to the internal ACU software calculating the turnaround trajectory using a cubic spline interpolation, we chose to verify satisfactory turnaround behavior during our offline analysis. An example of this turnaround behavior is shown in Figure \ref{fig:progtrack_currs}. In addition to the positional measurements, we calculate the trajectory velocity and acceleration using the \texttt{numpy.gradient} method. We conclude that the acceleration is sufficiently low throughout the turnarounds, and the motion is sufficiently smooth.

\begin{figure} [h!]
\begin{center}
\begin{tabular}{c}
\includegraphics[width=0.7\linewidth]{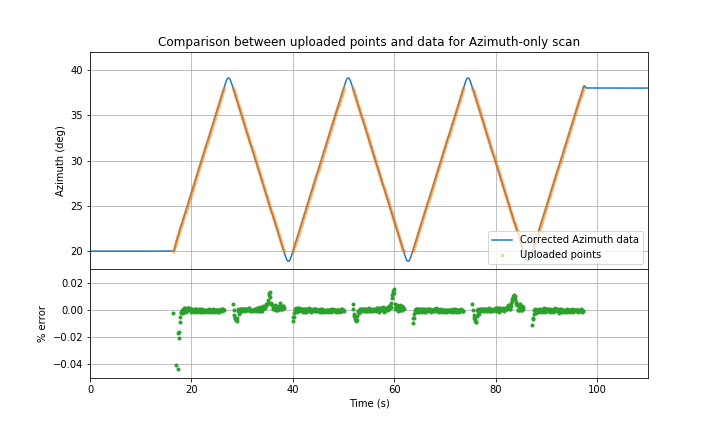}
\end{tabular}
\end{center}
\caption[example]
{ \label{fig:upload_matching}
We compare the trajectory values uploaded to the ACU to the encoder-measured trajectory during a constant velocity scan. The gaps between the last uploaded point before a turnaround and the next point are left when we calculate the scan trajectory to upload; we do not upload points within the turnaround, and instead leave this to the ACU cubic spline interpolation. We find that, for sufficiently spaced turnarounds, the encoder-measured values match the uploaded values to a high degree of precision ($<0.02\%$ error) during the majority of the scan. The greatest degree of error occurs at the beginning of the scan, due to friction.
}
\end{figure}

The accuracy of scan commanding is of high importance for executing well-controlled scans. To assess the scan uploading functionality, we compare the uploaded scan values to the encoder-measured positions. We find that for scans with a sufficiently low turnaround acceleration and sufficiently high velocity, the encoder-measured positions match the uploaded positions with $<0.02\%$ error, well within specifications (Figure \ref{fig:upload_matching}). We note that for low velocities and high turnaround accelerations, the turnaround time as calculated by the Agent is not long enough for the platform to complete the turnaround, resulting in less accurate scanning.

\begin{figure} [h!]
\begin{center}
\begin{tabular}{c c} 
\includegraphics[width=0.45\linewidth]{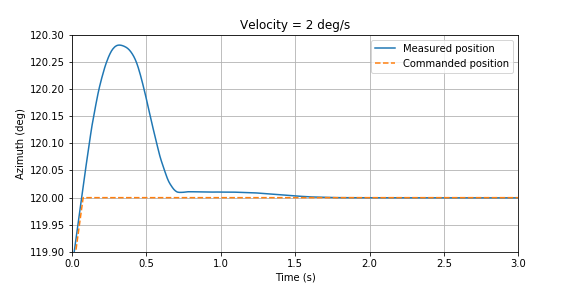}
&
\includegraphics[width=0.45\linewidth]{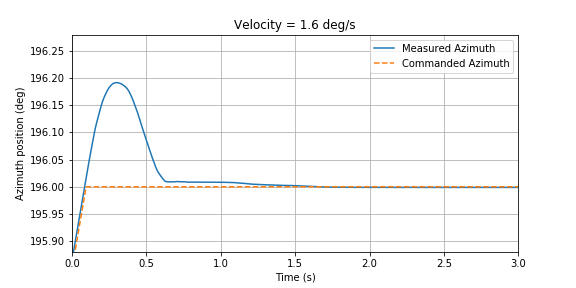}
\\
\includegraphics[width=0.45\linewidth]{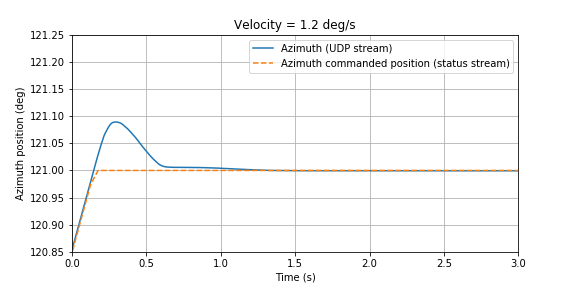}
&
\includegraphics[width=0.45\linewidth]{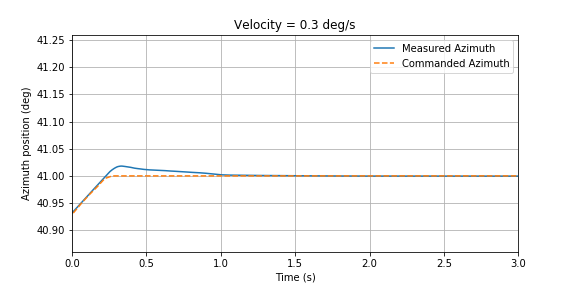}
\end{tabular}
\end{center}
\caption[example]
{ \label{fig:scansettles}
Examples of Azimuth-only scan completions with different scan velocities. We see that when completing scans with higher constant velocities, the SAT platform takes slightly longer to settle at the end positions. For the highest-velocity (2 degrees/second) scans measured in this manner, the platform takes about 1.5 seconds to settle at the final position; for the lowest velocity (0.1-0.3 degrees/second) measured, the platform takes about 1.0 second to settle.}
\end{figure}

The Agent uses rounding, with user-specified precision, to determine whether scans have completed their paths. However, if the rounding precision of the Agent is greater than the precision of the scan trajectory as determined by the ACU, the Agent may not complete the scan Task due to a mismatch of position values; if the rounding precision of the Agent is too low, the scan may complete too soon. To better determine the ends of scans with the Agent, we conducted an analysis of the stopping times and displacement from commanded scan ends for scans at several velocities (Figure \ref{fig:scansettles}). This comparison will allow us to assess how long the Agent should wait for the position to settle before completing a scan Task, ranging from $\sim$1 second to $\sim$1.5 seconds. The velocity-dependent settling time and analysis of appropriate rounding precision will be used to improve the end conditions for the scan Task.

\begin{figure} [h!]
\begin{center}
\begin{tabular}{c c} 
\includegraphics[width=0.45\linewidth]{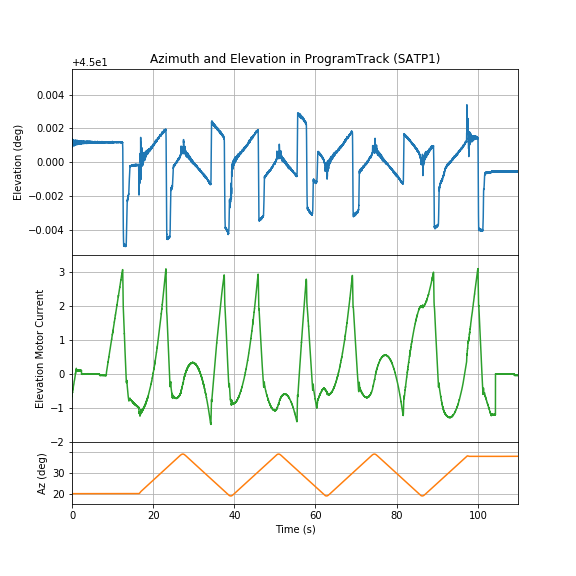}
&
\includegraphics[width=0.45\linewidth]{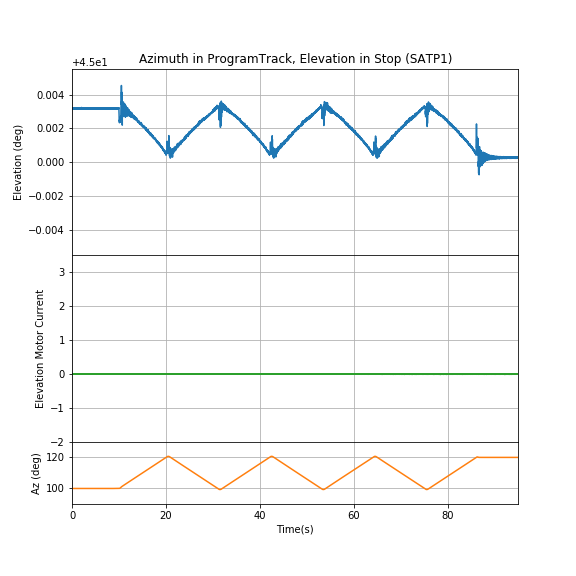}
\end{tabular}
\end{center}
\caption[example]
{ \label{fig:elbob}
We assess the position accuracy and predictability when executing Azimuth-only scans with the Elevation axis in the uploaded track mode (left) and stop mode (right). In both cases, the Elevation position is set to 45$\degree$, but exhibits some motion away from this set position. We can see that the Elevation axis exhibits random behavior and a high amount of current through the motor in the uploaded track mode. In stop mode, the Elevation exhibits a predictable bobble, with significantly less current through the motor.
}
\end{figure}

We additionally tested the Elevation axis behavior during Azimuth-only scans. Because we can perform Azimuth-only scans with the Elevation axis set to either the uploaded scan mode or stop mode, we tested Azimuth scans with both of these mode options for the Elevation axis to determine what the best software practice is for scans. Examples of the results of these tests are shown in Figure \ref{fig:elbob}. We determined that when the Elevation mode is set to the upload scan mode, the Elevation axis exhibits a random variation and draws current through its motor. When the Elevation mode is kept in stop mode during an Azimuth scan, the Elevation motors draw almost no current, and exhibits a small, periodic bobble. We concluded that, because the bobble is predictable and repeatable and the motor current is limited, the best practice is to set the Elevation axis to stop mode when performing Azimuth-only scans.

%%%%%%%%%%%%%%%%%%%%%%%%%%%%%%%%%%%%%%%%%%%%%

\section{Future Work} \label{sec:future}

\subsection{SAT In-Situ Testing} \label{subsec:sat_chile}
Prior to installing the SAT receivers, we will run acceptance tests again with only the platforms at the site in Chile. SAT platform testing is expected to begin in October 2022. We expect this testing to be similar in style to the SAT factory testing, with VA-administered data acquisition and command testing followed by \ocs{} command testing. We plan to complete a similar series of scan tests to assess the ACU and platform performance at the site, and to inform future work on the pointing model and scan conditions. We will additionally have the opportunity to test the functionality of the \texttt{generate\_scan} Process with the SAT platforms without the constraints of the factory environment.

\subsection{LAT Factory Testing and Deployment} \label{subsec:lat}
\soaculib{} and the ACU Agent are written with implementation for the SATs and the LAT in mind. We anticipate some updates to the Agent and \soaculib{} with the implementation of the LAT ACU software, including the added functionality of commanding the 3rd Axis during scans, a new set of Status fields, and potentially some differences in naming conventions within \soaculib{}. However, we expect these updates to be minor.

%%%%%%%%%%%%%%%%%%%%%%%%%%%%%%%%%%%%%%%%%%%%%%%%%%

\section{Summary} \label{sec:summary}
We have presented an overview of the software needed to integrate the telescope ACUs into the \ocs{} framework for operation within the larger observatory. The ACU Agent provides an interface between the ACU software and the \ocs{} framework, including operations that support data acquisition, live monitoring, and commanding. We additionally built a software simulator of the ACU, which is currently being used for full observatory end-to-end software testing.

The ACU Agent has been implemented with the SAT platforms in the factory setting during the platform acceptance testing. During this testing, we determined that the trajectory profiles of scans meet our requirements, and that these trajectories are highly accurate to the commanded scan profiles. This is important for our ability to complete the scan patterns needed to make observations to achieve our science goals. We additionally determined software best practices for Azimuth-only scans.

We expect to continue evaluating these best practices and software implementation as we begin system testing with the first SAT at the site in Fall 2022. We additionally expect to make minor adaptations to the ACU Agent for the LAT software in the near future.

\appendix
\section{SAT ACU Agent Data Fields}\label{sec:appendix}

\subsection{SAT \texttt{broadcast} Fields}

\begin{longtable}{|p{0.3\linewidth}|p{0.6\linewidth}|}
\caption{Fields published by the ACU Agent \texttt{broadcast} Process when configured with a SAT.}
\label{tab:bcast_fields}\\
\hline
\textbf{Field name} & \textbf{Field description}\\
\hline
\texttt{Time} & The timestamp, as seconds since the epoch. This value is calculated from the ACU's internal timestamp and a \texttt{datetime} year timestamp.\\
\hline
\texttt{Raw\_Azimuth} & The integrated encoder values on the Azimuth axis, before any corrections are applied.\\
\hline
\texttt{Raw\_Elevation} & The integrated encoder values on the Elevation axis, before any corrections are applied.\\
\hline
\texttt{Raw\_Boresight} & The integrated encoder values on the Boresight axis, before any corrections are applied.\\
\hline
\texttt{Corrected\_Azimuth} & The integrated encoder values on the Azimuth axis, with pointing and refraction parameter corrections applied.\\
\hline
\texttt{Corrected\_Elevation} & The integrated encoder values on the Elevation axis, with pointing and refraction parameter corrections applied.\\
\hline
\texttt{Corrected\_Boresight} & The integrated encoder values on the Boresight axis, with pointing and refraction parameter corrections applied.\\
\hline
\texttt{Azimuth\_Current\_1} & The current drawn by Motor 1 on the Azimuth axis.\\
\hline
\texttt{Azimuth\_Current\_2} & The current drawn by Motor 2 on the Azimuth axis.\\
\hline
\texttt{Elevation\_Current\_1} & The current drawn by Motor 1 on the Elevation axis.\\
\hline
\texttt{Boresight\_Current\_1} & The current drawn by Motor 1 on the Boresight axis.\\
\hline
\texttt{Boresight\_Current\_2} & The current drawn by Motor 2 on the Boresight axis.\\
\hline

\end{longtable}

\subsection{SAT \texttt{monitor} Fields}

\begin{center}       
\begin{longtable}{|p{0.38\linewidth}|p{0.52\linewidth}|}
\caption{Fields published by the ACU Agent \texttt{monitor} Process when configured with a SAT.} 
\label{tab:monitor_fields_summary}\endfirsthead
\hline
\textbf{Field name} & \textbf{Field description}\endhead
\hline
\textbf{Field name} & \textbf{Field description}\\
\hline
\texttt{Time} & The timestamp from the ACU, recorded as the decimal day of the year.\\
\hline
\texttt{Year} & The year, as timestamped by the ACU.\\
\hline
\texttt{ctime} & The timestamp, as seconds since the epoch. This value is calculated from the ACU's internal \texttt{Time} and \texttt{Year} timestamps.\\
\hline
\texttt{Azimuth\_mode} & The motion mode set on the Azimuth axis.\\
\hline
\texttt{Azimuth\_current\_position} & The encoder position on the Azimuth axis, after any corrections (degrees).\\
\hline
\texttt{Azimuth\_current\_velocity} & The measured velocity on the Azimuth axis (degrees/second).\\
\hline
\texttt{Elevation\_mode} & The motion mode set on the Elevation axis.\\
\hline
\texttt{Elevation\_current\_position} & The encoder position on the Elevation axis, after any corrections (degrees).\\
\hline
\texttt{Elevation\_current\_velocity} & The measured velocity on the Elevation axis (degrees/second).\\
\hline
\texttt{Boresight\_mode} & The motion mode set on the Boresight axis.\\
\hline
\texttt{Boresight\_current\_position} & The encoder position on the Boresight axis, after any corrections (degrees).\\
\hline
\texttt{Boresight\_current\_velocity} & The measured velocity on the Boresight axis (degrees/second).\\
\hline
\texttt{Free\_upload\_positions} & The number of unused possible upload points in the scan upload stack.\\
\hline
\texttt{Azimuth\_avg\_position\_error} & The average difference between achieved and commanded position on the Azimuth axis.\\
\hline
\texttt{Azimuth\_peak\_position\_error} & The peak difference between achieved and commanded position on the Azimuth axis.\\
\hline
\texttt{Elevation\_avg\_position\_error} & The average difference between achieved and commanded position on the Elevation axis.\\
\hline
\texttt{Elevation\_peak\_position\_error} & The peak difference between achieved and commanded position on the Elevation axis.\\
\hline
\texttt{AzCCW\_SWprelimit} & ACU software pre-limit for motion in the Azimuth counter-clockwise direction is triggered (boolean).\\
\hline
\texttt{AzCCW\_SWlimit\_operating} & ACU software limit for motion in the Azimuth counter-clockwise direction is triggered (boolean). The ACU software will not command positions beyond this limit.\\
\hline
\texttt{AzCCW\_HWprelimit} & Hardware pre-limit for motion in the Azimuth counter-clockwise direction is triggered (boolean).\\
\hline
\texttt{AzCCW\_HWlimit\_operating} & Operating limit for motion in the Azimuth counter-clockwise direction is triggered (boolean).\\
\hline
\texttt{AzCCW\_HWlimit\_emergency} & Emergency limit for motion in the Azimuth counter-clockwise direction is triggered (boolean).\\
\hline
\texttt{AzCCW\_HWlimit\_2ndEmergency} & Second emergency limit for motion in the Azimuth counter-clockwise direction is triggered (boolean).\\
\hline
\texttt{AzCW\_SWprelimit} & ACU software pre-limit for motion in the Azimuth clockwise direction is triggered (boolean).\\
\hline
\texttt{AzCW\_SWlimit\_operating} & ACU software limit for motion in the Azimuth clockwise direction is triggered (boolean). The ACU software will not command positions beyond this limit.\\
\hline
\texttt{AzCW\_HWprelimit} & Hardware pre-limit for motion in the Azimuth clockwise direction is triggered (boolean).\\
\hline
\texttt{AzCW\_HWlimit\_operating} & Operating limit for motion in the Azimuth clockwise direction is triggered (boolean).\\
\hline
\texttt{AzCW\_HWlimit\_emergency} & Emergency limit for motion in the Azimuth clockwise direction is triggered (boolean).\\
\hline
\texttt{AzCW\_HWlimit\_2ndEmergency} & Second emergency limit for motion in the Azimuth clockwise direction is triggered (boolean).\\
\hline
\texttt{ElDown\_SWprelimit} & ACU software pre-limit for motion in the Elevation downward direction is triggered (boolean).\\
\hline
\texttt{ElDown\_SWlimit\_operating} & ACU software limit for motion in the Elevation downward direction is triggered (boolean). The ACU software will not command positions beyond this limit.\\
\hline
\texttt{ElDown\_HWprelimit} & Hardware pre-limit for motion in the Elevation downward direction is triggered (boolean).\\
\hline
\texttt{ElDown\_HWlimit\_operating} & Operating limit for motion in the Elevation downward direction is triggered (boolean).\\
\hline
\texttt{ElDown\_HWlimit\_emergency} & Emergency limit for motion in the Elevation downward direction is triggered (boolean).\\
\hline
\texttt{ElDown\_HWlimit\_shieldOFF\_operating} & Operating limit for motion in the Elevation downward direction with the co-moving shield off is triggered (boolean).\\
\hline
\texttt{ElDown\_HWlimit\_shieldOFF\_emergency} & Emergency limit for motion in the Elevation downward direction with the co-moving shield off is triggered (boolean).\\
\hline
\texttt{ElUp\_SWprelimit} & ACU software pre-limit for motion in the Elevation upward direction is triggered (boolean).\\
\hline
\texttt{ElUp\_SWlimit\_operating} & ACU software limit for motion in the Elevation upward direction is triggered (boolean). The ACU software will not command positions beyond this limit.\\
\hline
\texttt{ElUp\_HWprelimit} & Hardware pre-limit for motion in the Elevation upward direction is triggered (boolean).\\
\hline
\texttt{ElUp\_HWlimit\_operating} & Operating limit for motion in the Elevation upward direction is triggered (boolean).\\
\hline
\texttt{ElUp\_HWlimit\_emergency} & Emergency limit for motion in the Elevation upward direction is triggered (boolean).\\
\hline
\texttt{BsCCW\_SWprelimit} & ACU software pre-limit for motion in the Boresight counter-clockwise direction is triggered (boolean).\\
\hline
\texttt{BsCCW\_SWlimit\_operating} & ACU software limit for motion in the Boresight counter-clockwise direction is triggered (boolean). The ACU software will not command positions beyond this limit.\\
\hline
\texttt{BsCCW\_HWprelimit} & Hardware pre-limit for motion in the Boresight counter-clockwise direction is triggered (boolean).\\
\hline
\texttt{BsCCW\_HWlimit\_operating} & Operating limit for motion in the Boresight counter-clockwise direction is triggered (boolean).\\
\hline
\texttt{BsCCW\_HWlimit\_emergency} & Emergency limit for motion in the Boresight counter-clockwise direction is triggered (boolean).\\
\hline
\texttt{BsCW\_SWprelimit} & ACU software pre-limit for motion in the Boresight clockwise direction is triggered (boolean).\\
\hline
\texttt{BsCW\_SWlimit\_operating} & ACU software limit for motion in the Boresight clockwise direction is triggered (boolean). The ACU software will not command positions beyond this limit.\\
\hline
\texttt{BsCW\_HWprelimit} & Hardware pre-limit for motion in the Boresight clockwise direction is triggered (boolean).\\
\hline
\texttt{BsCW\_HWlimit\_operating} & Operating limit for motion in the Boresight clockwise direction is triggered (boolean).\\
\hline
\texttt{BsCW\_HWlimit\_emergency} & Emergency limit for motion in the Boresight clockwise direction is triggered (boolean).\\
\hline
\texttt{Azimuth\_summary\_fault} & Azimuth Mode is changed to Stop by any failure (boolean).\\
\hline
\texttt{Azimuth\_motion\_error} & Azimuth axis motion not detected although commanded to do so (boolean).\\
\hline
\texttt{Azimuth\_motor1\_overtemp} & Azimuth motor 1 temperature over limit (boolean).\\
\hline
\texttt{Azimuth\_motor2\_overtemp} & Azimuth motor 2 temperature over limit (boolean).\\
\hline
\texttt{Azimuth\_overspeed} & Low level Azimuth speed over limit (boolean).\\
\hline
\texttt{Azimuth\_resistor1\_overtemp} & Azimuth regeneration resistor 1 temperature over limit (boolean).\\
\hline
\texttt{Azimuth\_resistor2\_overtemp} & Azimuth regeneration resistor 2 temperature over limit (boolean).\\
\hline
\texttt{Azimuth\_motor1\_overcurrent} & Azimuth motor 1 current over limit (boolean).\\
\hline
\texttt{Azimuth\_motor2\_overcurrent} & Azimuth motor 2 current over limit (boolean).\\
\hline
\texttt{Elevation\_summary\_fault} & Elevation mode is changed to Stop by any failure (boolean).\\
\hline
\texttt{Elevation\_motion\_error} & Elevation axis motion not detected although commanded to do so (boolean).\\
\hline
\texttt{Elevation\_motor1\_overtemp} & Elevation motor 1 temperature over limit (boolean).\\
\hline
\texttt{Elevation\_overspeed} & Low level Elevation speed over limit (boolean).\\
\hline
\texttt{Elevation\_resistor1\_overtemp} & Elevation regeneration resistor 1 temperature over limit (boolean).\\
\hline
\texttt{Elevation\_motor1\_overcurrent} & Elevation motor 1 current over limit (boolean).\\
\hline
\texttt{Boresight\_summary\_fault} & Boresight Mode is changed to Stop by any failure (boolean).\\
\hline
\texttt{Boresight\_motion\_error} & Boresight axis motion not detected although commanded to do so (boolean).\\
\hline
\texttt{Boresight\_motor1\_overtemp} & Boresight motor 1 temperature over limit (boolean).\\
\hline
\texttt{Boresight\_motor2\_overtemp} & Boresight motor 2 temperature over limit (boolean).\\
\hline
\texttt{Boresight\_overspeed} & Low level Boresight speed over limit (boolean).\\
\hline
\texttt{Boresight\_resistor1\_overtemp} & Boresight regeneration resistor 1 temperature over limit (boolean).\\
\hline
\texttt{Boresight\_resistor2\_overtemp} & Boresight regeneration resistor 2 temperature over limit (boolean).\\
\hline
\texttt{Boresight\_motor1\_overcurrent} & Boresight motor 1 current over limit (boolean).\\
\hline
\texttt{Boresight\_motor2\_overcurrent} & Boresight motor 2 current over limit (boolean).\\
\hline
\texttt{Azimuth\_oscillation\_warning} & Azimuth oscillation warning detected (boolean).\\
\hline
\texttt{Elevation\_oscillation\_warning} & Elevation oscillation warning detected (boolean).\\
\hline
\texttt{Boresight\_oscillation\_warning} & Boresight oscillation warning detected (boolean).\\
\hline
\texttt{Azimuth\_servo\_failure} & Low level failure on Azimuth axis detected (boolean).\\
\hline
\texttt{Azimuth\_brake1\_failure} & Unexpected current consumption by Azimuth brake 1 (boolean).\\
\hline
\texttt{Azimuth\_brake2\_failure} & Unexpected current consumption by Azimuth brake 2 (boolean).\\
\hline
\texttt{Azimuth\_breaker\_failure} & Any Azimuth circuit breaker has been tripped (boolean).\\
\hline
\texttt{Az\_CANbus\_amp1\_comms\_failure} & Fault detected in Azimuth CAN bus amplifier 1 communication (boolean).\\
\hline
\texttt{Az\_CANbus\_amp2\_comms\_failure} & Fault detected in Azimuth CAN bus amplifier 2 communication (boolean).\\
\hline
\texttt{Azimuth\_encoder\_failure} & Fault detected in Azimuth encoder (boolean).\\
\hline
\texttt{Azimuth\_tacho\_failure} & Difference in Azimuth motor speeds over limit (boolean).\\
\hline
\texttt{Elevation\_servo\_failure} & Low level failure on Elevation axis detected (boolean).\\
\hline
\texttt{Elevation\_brake1\_failure} & Unexpected current consumption by Elevation brake 1 (boolean).\\
\hline
\texttt{Elevation\_breaker\_failure} & Any Elevation circuit breaker has been tripped (boolean).\\
\hline
\texttt{El\_CANbus\_amp1\_comms\_failure} & Fault detected in Elevation CAN bus amplifier 1 communication (boolean).\\
\hline
\texttt{Elevation\_encoder\_failure} & Fault detected in Elevation encoder (boolean).\\
\hline
\texttt{Boresight\_servo\_failure} & Low level failure on Boresight axis detected (boolean).\\
\hline
\texttt{Boresight\_brake1\_failure} & Unexpected current consumption by Boresight brake 1 (boolean).\\
\hline
\texttt{Boresight\_brake2\_failure} & Unexpected current consumption by Boresight brake 2 (boolean).\\
\hline
\texttt{Boresight\_breaker\_failure} & Any Boresight circuit breaker has been tripped (boolean).\\
\hline
\texttt{Bs\_CANbus\_amp1\_comms\_failure} & Fault detected in Boresight CAN bus amplifier 1 communication (boolean).\\
\hline
\texttt{Bs\_CANbus\_amp2\_comms\_failure} & Fault detected in Boresight CAN bus amplifier 2 communication (boolean).\\
\hline
\texttt{Boresight\_encoder\_failure} & Fault detected in Boresight encoder (boolean).\\
\hline
\texttt{Boresight\_tacho\_failure} & Difference in Boresight motor speeds over limit (boolean).\\
\hline
\texttt{Azimuth\_computer\_disabled} & ACU is not in control of the Azimuth axis (boolean).\\
\hline
\texttt{Azimuth\_axis\_stop} & Azimuth axis is in Stop Mode (boolean).\\
\hline
\texttt{Azimuth\_brakes\_released} & All Azimuth brakes are released (boolean).\\
\hline
\texttt{Azimuth\_stop\_LCP} & Push-button Azimuth Stop at the Local Control Panel (LCP) is activated (boolean).\\
\hline
\texttt{Azimuth\_power\_on} & Azimuth drive power is turned on (boolean).\\
\hline
\texttt{Azimuth\_AUX1\_mode\_selected} & Auxiliary mode 1 on the Azimuth axis is selected (boolean).\\
\hline
\texttt{Azimuth\_AUX2\_mode\_selected} & Auxiliary mode 2 on the Azimuth axis is selected (boolean).\\
\hline
\texttt{Azimuth\_immobile} & Azimuth axis is not moving (boolean).\\
\hline
\texttt{Elevation\_computer\_disabled} & ACU is not in control of the Elevation axis (boolean).\\
\hline
\texttt{Elevation\_axis\_stop} & Elevation axis is in Stop Mode (boolean).\\
\hline
\texttt{Elevation\_brakes\_released} & All Elevation brakes are released (boolean).\\
\hline
\texttt{Elevation\_stop\_LCP} & Push-button Elevation Stop at the LCP is activated (boolean).\\
\hline
\texttt{Elevation\_power\_on} & Elevation drive power is turned on (boolean).\\
\hline
\texttt{Elevation\_immobile} & Elevation axis is not moving (boolean).\\
\hline
\texttt{Boresight\_computer\_disabled} & ACU is not in control of the Boresight axis (boolean).\\
\hline
\texttt{Boresight\_axis\_stop} & Boresight axis is in Stop Mode (boolean).\\
\hline
\texttt{Boresight\_brakes\_released} & All Boresight brakes are released (boolean).\\
\hline
\texttt{Boresight\_stop\_LCP} & Push-button Boresight Stop at the LCP is activated (boolean).\\
\hline
\texttt{Boresight\_power\_on} & Boresight drive power is turned on (boolean).\\
\hline
\texttt{Boresight\_AUX1\_mode\_selected} & Auxiliary mode 1 on the Boresight axis is selected (boolean).\\
\hline
\texttt{Boresight\_AUX2\_mode\_selected} & Auxiliary mode 2 on the Boresight axis is selected (boolean).\\
\hline
\texttt{Boresight\_immobile} & Boresight axis is not moving (boolean).\\
\hline
\texttt{Azimuth\_oscillation\_alarm} & Azimuth oscillation alarm detected (boolean).\\
\hline
\texttt{Elevation\_oscillation\_alarm} & Elevation oscillation alarm detected (boolean).\\
\hline
\texttt{Boresight\_oscillation\_alarm} & Boresight oscillation alarm detected (boolean).\\
\hline
\texttt{Azimuth\_commanded\_position} & Input to the control loop on the Azimuth axis. Value type is \texttt{float} or \texttt{None}.\\
\hline
\texttt{Elevation\_commanded\_position} & Input to the control loop on the Elevation axis. Value type is \texttt{float} or \texttt{None}.\\
\hline
\texttt{Boresight\_commanded\_position} & Input to the control loop on the Boresight axis. Value type is \texttt{float} or \texttt{None}.\\
\hline
\texttt{General\_summary\_fault} & Mode is changed to Stop by the ACU for all axes due to any failure (boolean).\\
\hline
\texttt{Power\_failure\_Latched} & Short break power is missing and ACU is latched (boolean).\\
\hline
\texttt{Power\_failure\_24V} & Internal 24 V power is missing (boolean).\\
\hline
\texttt{General\_breaker\_failure} & Any breaker on any axis is tripped (boolean).\\
\hline
\texttt{Power\_failure\_NotLatched} & Short break power is missing and ACU is not latched (boolean).\\
\hline
\texttt{Cabinet\_overtemp} & Measured cabinet temperature is over limit (boolean).\\
\hline
\texttt{Ambient\_temp\_TooLow} & Ambient temperature is too low, operation is inhibited (boolean).\\
\hline
\texttt{PLC\_interface\_error} & Programmable logic conroller (PLC) detected an interface error (boolean).\\
\hline
\texttt{ACU\_fan\_failure} & ACU detects a fan failure (boolean).\\
\hline
\texttt{Cabinet\_undertemp} & Cabinet temperature below limit (boolean).\\
\hline
\texttt{Time\_sync\_error} & PTP time synchronization failure detected (boolean).\\
\hline
\texttt{PLC\_comms\_error} & PLC communication error detected\\
\hline
\texttt{EStop\_servo\_drive\_cabinet} & E-Stop triggered at the servo drive cabinet\\
\hline
\texttt{EStop\_service\_pole} & E-Stop triggered at the service pole\\
\hline
\texttt{EStop\_Az\_movable} & E-Stop triggered with Azimuth axis movable (boolean).\\
\hline
\texttt{Key\_switch\_bypass\_emergency\_limit} & Key switch engaged to bypass an emergency limit (boolean).\\
\hline
\texttt{PCU\_operation} & PCU is active (boolean).\\
\hline
\texttt{Safe\_mode} & Safe mode activated (boolean).\\
\hline
\texttt{Lightning\_protection\_surge\_arresters} & Any lightning protection surge arrestor tripped (boolean).\\
\hline
\texttt{CoMoving\_shield\_off} & Co-moving shield is off (boolean).\\
\hline
\texttt{Remote\_mode} & ACU is in remote mode, not local mode (boolean).\\
\hline
\end{longtable}
\end{center}
\acknowledgements
This work is supported in part by a grant from the Simons Foundation (Award \#457687, B.K.).

\bibliographystyle{spiebib}
\bibliography{refs.bib}

\end{document}